\documentclass[twocolumn,pra,longbibliography]{revtex4-1}
\usepackage{graphicx}
\usepackage{float}
\usepackage{dcolumn}
\usepackage{bm}
\usepackage{color}
\usepackage{amsmath}
\usepackage{amssymb}
\usepackage{amsfonts}
\usepackage{esint}
\usepackage{times}
\usepackage{xcolor}
\usepackage{phaistos}
\usepackage{braket}
\usepackage{comment}
\usepackage{multirow}
\usepackage{diagbox} %% For the diagonal box
\usepackage{tabu}

\usepackage[colorlinks,linkcolor=cyan,anchorcolor=cyan,urlcolor=cyan,citecolor=cyan]{hyperref}

\begin{document}

\title{Microwave-activated two-qubit gates for fixed-coupling and fixed-frequency transmon qubits}

\author{Ling Jiang$^{1,3}$}
\author{Peng Xu$^{2}$}\email{pengxu@njupt.edu.cn}
\author{Shengjun Wu$^{3}$}\email{sjwu@nju.edu.cn}
\author{Jian-An Sun$^{1}$}
\author{Fu-Quan Dou$^{1}$}\email{doufq@nwnu.edu.cn}
\affiliation{$^1$ College of Physics and Electronic Engineering, Northwest Normal University, Lanzhou, 730070, China\\
$^2$ Institute of Quantum Information and Technology, Nanjing University of Posts and Telecommunications, Nanjing 210003, China\\
$^3$ National Laboratory of Solid State Microstructures and School of Physics, Collaborative Innovation Center
of Advanced Microstructures, Nanjing University, Nanjing 210093, China}

\date{\today}

\begin{abstract}
  
All-microwave control of fixed-frequency superconducting quantum systems offers the potential to reduce control circuit complexity and increase system coherence. Nevertheless, due to the limited control flexibility in qubit parameters, one has to address several issues, such as quantum crosstalk and frequency crowding, for scaling up qubit architecture with non-tunable elements. This study proposes a microwave-activated two-qubit gate scheme for two fixed-frequency transmon qubits coupled via a fixed-frequency transmon coupler. The protocol relies on applying a microwave pulse exclusively to the coupler, enabling the implementation of a controlled-Z (CZ) gate. We show that the gate fidelity exceeding 0.999 can be achieved within 150 ns, excluding decoherence effects. Moreover, we also show that leakage from the computational subspace to non-computational states can also be effectively suppressed.

\end{abstract}

\maketitle

%\pacs{03.67.Lx, 03.67.Bg, 85.25.Cp}

%$\textbf{OCIS  codes:}$ (020.5580) Quantum electrodynamics; (060.5565) Quantum communications; (190.0190) Nonlinear optics; (270.0270) Quantum optics.

%Keywords: superconducting quantum processor, superconducting qutrit, quantum gate

\section{Introduction}

Quantum information science, which integrates quantum mechanics with information science, has entered a new stage, particularly with the advent of experimental advancements that have brought us into the noisy intermediate-scale quantum (NISQ) era \cite{JPreskillQuantum2018, FAruteNature5052019, AOmranScience3652019, CSongScience3652019, ZYanScience3642019}. The overarching aim is to create a programmable quantum computer capable of solving complex problems beyond classical computers' capabilities. Based on a wide variety of physical systems, significant efforts are being made to discover potential applications for quantum computing. Superconducting circuits are particularly promising for constructing quantum computers among the various platforms due to their advantages, such as ease of control and integration and expansion, especially within the NISQ era. However, realizing practical quantum advantage with NISQ devices is critically dependent on the precision of quantum operations, with two-qubit operations being the main limiting factor, a challenge that typically escalates as more qubits and control lines are incorporated.

Over the past decades, numerous protocols have been proposed and experimentally demonstrated for implementing two-qubit gates in coupled superconducting qubit systems. One approach involves dynamically tuning components of the superconducting circuit, such as the qubit frequency \cite{RJSchoelkopfNature4602009, MRGellerPRA902014, JMMartinisNature5082014, MARolPRL1231205022019, LDiCarloPRL1262205022021}, as well as the coupling strength \cite{JMMartinisPRL1132014, JMGambettaPRApp62016, FYanPRApp102018, PSMundadaPRApp122019, MCCollodoPRL1252405022020, YuanXuPRL1252405032020, HGotoPRApp180340382022, DLCampbellPRApp190640432023} between them. While the approach with dynamic-frequency tuning can enable fast-speed gates, it may also introduce noise and crosstalk channels, leading to decoherence and reduced gate fidelity. Alternatively, microwave pulses can be used to implement universal control of qubits including both single- and two-qubit gates \cite{GSParaoanuPRB74140504R2006, PCdeGrootNP67632010, CRigettiPRB812010, JMChowPRL1072011, TNohSR82018, SPolettoPRL1092012, JMChowNJP152013, HPaikPRL1172016, SPPremaratnePRA992019, SKrinnerPRApp0440392020, BKMitchell1272021, KXWeiPRL1290605012022}. This method typically targets a fixed-frequency, fixed-coupling superconducting qubit architecture. Unlike frequency-tunable gate schemes, microwave-activated gates for fixed-frequency qubits offer longer coherence times and simpler control circuit requirements. Due to these advantages, fixed-frequency transmons \cite{JKochPRA762007} have become a leading choice for building quantum computers.

Despite the advantages mentioned above, fixed-coupling, there are also many challenges for scaling up this qubit architecture with non-tunable elements, perhaps the greatest one is the quantum crosstalk, i.e., static ZZ coupling \cite{RJSchoelkopfNature4602009, PSMundadaPRApp122019, SPForsarXiv2408154022024}, due to the fixed qubit couplings. This coupling can degrade the performance of quantum gates and induce crosstalk between qubits, thereby complicating error correction and quantum operations. Although idling errors and quantum crosstalk due to the always-on ZZ coupling can be effectively mitigated through various circuit designs \cite{PZhaoXuPRL1252020, JKuPRL1252005042020, AKandalaPRL1271305012021, PengZhaoPRAp0240372021}, the implementation of these schemes involves significant complexity and challenges. Meanwhile, in this qubit architecture with non-tunable components, two-qubit gate operations are generally based on employing microwave drives applied to the qubits \cite{JMChowPRL1072011, SKrinnerPRApp0440392020, BKMitchell1272021, KXWeiPRL1290605012022, KXWeiPRXQuantum52024}. However, in a large qubit lattice, the qubit drive may impact or be constrained by neighboring qubits, thus complicating the qubit control in the architecture and potentially limiting its scalability. The above two issues can be largely addressed by (1) using a multi-path coupler for suppressing the ZZ crosstalk \cite{AKandalaPRL1271305012021, PengZhaoPRAp0240372021} and (2) employing gate schemes with the drive solely applied to the coupler. This strategy offers the advantage of better scalability for achieving high-fidelity quit control in large-scale qubit systems. A direct implementation, which combines the multi-path coupler for ZZ-suppression and two-qubit CZ gates with a microwave-driven coupler has been demonstrated in Ref. \cite{YNakamuraPRL2606012023}. However, (1) the suppression of ZZ interactions in the gate architecture with a multi-path coupler highly hinges on the precision in setting the qubit and coupling parameters, i.e., operating in the straddling regime \cite{AKandalaPRL1271305012021, PengZhaoPRAp0240372021}; (2) the longer gate time and the need for strong drive strength in Ref. \cite{YNakamuraPRL2606012023} are not ideal for achieving high-fidelity gates. These two figures make it challenging to maintain high-performance qubit control in large systems.

In this work, we propose a gate architecture where fixed-frequency transmon qubits are coupled via a non-tunable transmon coupler, and a CZ gate is realized by solely applying a microwave drive to the coupler. Moreover, to achieve ZZ suppression, here we consider the ABC-type frequency architecture \cite{PengXuarXiv24042024}, which incorporates three distinct frequencies for the qubits and the coupler, with the coupler's frequency situated between the frequencies of the two qubits. Compared to the gate architecture proposed in \cite{YNakamuraPRL2606012023}, our approach can achieve a two-qubit CZ gate fidelity of up to 0.999 within 150 ns using a drive pulse with moderate amplitude. Given the ABC-type architecture, the static ZZ coupling can be significantly suppressed. Moreover, optimizing the control pulse parameters can further reduce the gate time. Our gate scheme requires only a few megahertz of driving strength, which is far smaller than that used in Ref. \cite{YNakamuraPRL2606012023}, making it more achievable in practice and avoiding possible gate error sources, such as microwave crosstalk and chip heating, due to large-amplitude microwave drives. Importantly, the microwave manipulation of a fixed-frequency coupler significantly reduces the influence on or from neighboring qubits. Thus, this approach simplifies control in multi-qubit systems, enhances coherence, and facilitates the extension of two-qubit gates to larger quantum qubit systems.

The remainder of the paper is structured as follows: Section II introduces the superconducting circuit model, explores the ZZ coupling mechanism, and provides an overview of the methodology for the microwave-driven CZ gate scheme. Section III delves into the system dynamics under varying microwave pulse parameters during CZ gate operation. Section IV extends the proposed gate scheme to different system parameters and presents numerical simulations of the system's dynamics. Finally, the conclusions are summarized in Section V. Appendix A presents the analytical derivation of the effective ZZ coupling strength using perturbation theory.

\begin{figure}
\begin{center}
\includegraphics[width=8.80cm, height=4.450cm]{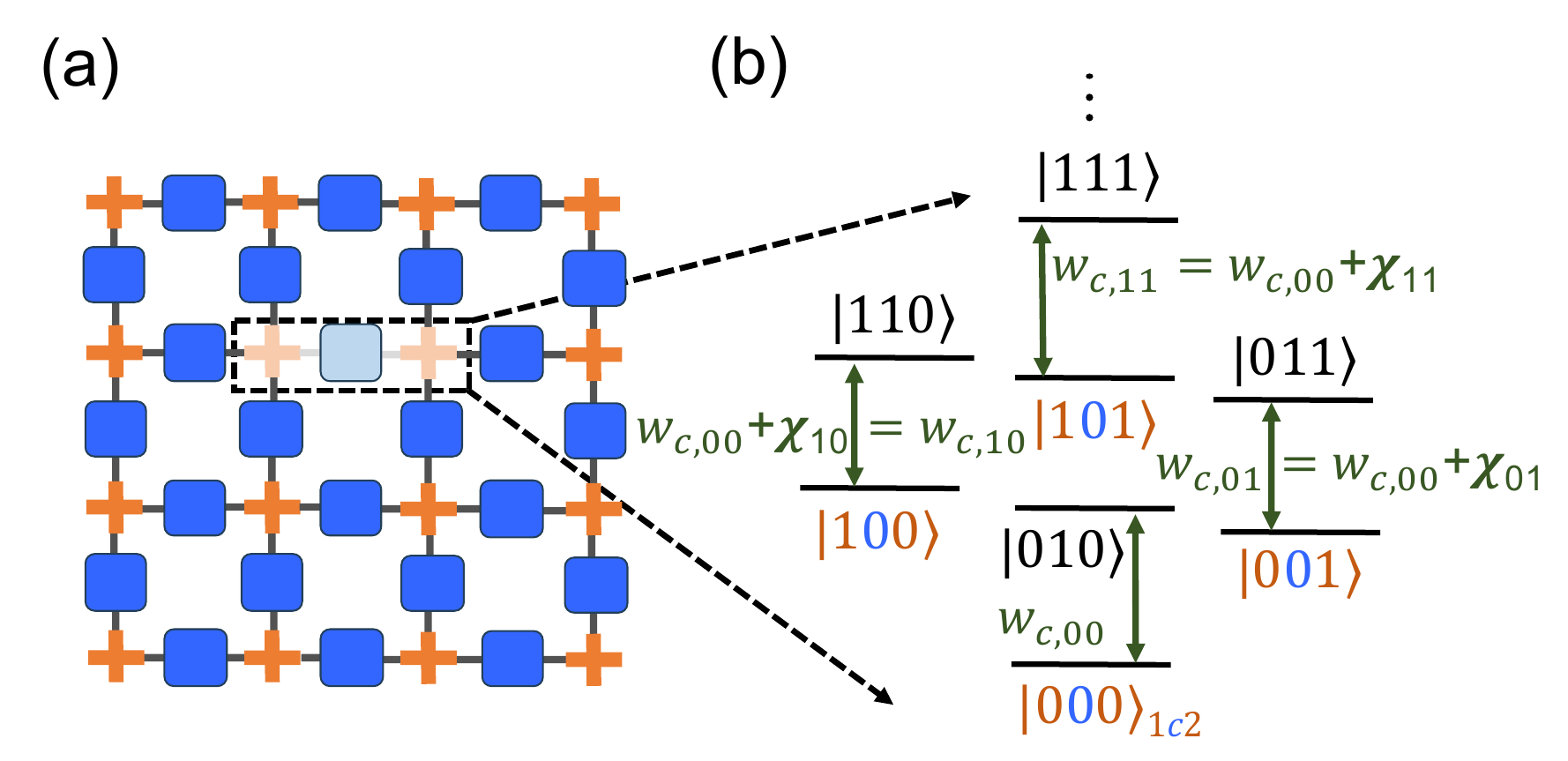}
\end{center}
\caption{(a) The layout of a two-dimensional coupling lattice, where light-orange crosses represent transmon qubits and light-blue squares denote transmon couplers. All qubits and couplers are capacitively coupled, as the grey lines indicate. The black dotted box highlights a pair of transmon qubits ($Q_{1,2}$) capacitively coupled to a common coupler qubit $Q_c$, driven by an external microwave pulse. (b) The system-level diagram within the black dotted box depicts the relevant qubit-coupler-qubit states and the transition frequencies between the four computational basis states and associated non-computational states. Here, $\omega_{c,00}$ denotes the frequency difference between the states $|000\rangle$ and $|010\rangle$, where the subscript $c$ indicates that the coupler is in the $|1\rangle$ state and $00$ signifies that the qubits are in the $|00\rangle$ state. The term $\chi_{mn}$ $(m,n \in {0,1})$ represents the difference between the frequencies $\omega_{c, mn}$ and $\omega_{c,00}$.}
\label{fig:one} 
\end{figure}

\section{Physical Model}

\subsection{ZZ coupling}

The superconducting circuit system comprises two transmon qubits $Q_{1(2)}$, capacitively coupled to a transmon coupler $Q_c$, as schematically depicted by the black-dotted box in Fig.~\ref{fig:one}(a). The transmon frequencies and qubit-coupler coupling strengths are fixed, with the qubits and coupler occupying three distinct frequency regimes. The coupler's frequency is placed between those of the two qubits. Under the rotating-wave approximation, the Hamiltonian of the qubit-coupler system is given by the following expression (with $\hbar = 1$), 
\begin{equation} 
\begin{aligned} H_s = \sum_{l} \left(\omega_l a^\dagger_l a_l + \frac{1}{2} \alpha_l a^\dagger_l a^\dagger_l a_l a_l\right) + \sum_{l \neq c}g_{lc}(a^\dagger_l a_c + a_l a^\dagger_c), 
\end{aligned} 
\end{equation} 
where $\omega_l$ and $\alpha_l$ $(l = 1, 2, c)$ represent the frequency and anharmonicity of each transmon, $a^\dagger_l$ and $a_l$ are the creation and annihilation operators, and $g_{lc}$ denotes the coupling strength between the transmon qubit $Q_l$ and the coupler $Q_c$. To account for the influence of higher energy levels, we consider up to the third excited state of each transmon \cite{FQDouPRA1070237252023, FQDouPRA1090624322024}. Figure~\ref{fig:one}(b) shows the transition frequencies between computational basis states and the related system energy levels.

To implement high-performance qubit control, one must address the primary parasitic interaction of ZZ coupling, which primarily results from interactions between the higher energy levels of qubits \cite{RJSchoelkopfNature4602009, FCWellstoodPRL912003}, in qubit architecture with non-tunable elements, such as the one studied in the present work. This issue is particularly significant in qubits with weak anharmonicity, such as transmon qubits \cite{JKochPRA762007}, where non-zero ZZ coupling is intrinsically present due to the qubits' energy level structure. The effect of static ZZ coupling extends beyond merely inducing the phase errors; when implementing gate operations, including both single- and two-qubit gates, the ZZ coupling can cause quantum crosstalk from neighboring spectator qubits, thus affecting gate fidelity in multi-qubit systems \cite{JMChowNJP152013, PSMundadaPRApp122019, DCMcKayPRL1222019, SKrinnerPRApp0240422020}. Therefore, mitigating the unwanted ZZ interaction is essential for enhancing gate fidelity in qubit architecture with fixed-frequency elements. Within the computational subspace, this interaction is characterized as ZZ coupling, with Z representing the Pauli operator $\sigma_z$, and is inherently present between any pair of qubits in the circuit, referred to as static ZZ interaction. The ZZ coupling strength $\zeta$ is given by,
 \begin{equation} 
 \begin{aligned} \zeta = E_{|\widetilde{101}\rangle} - E_{|\widetilde{001}\rangle} - E_{|\widetilde{100}\rangle} + E_{|\widetilde{000}\rangle}, 
 \end{aligned} 
 \end{equation} 
where $|\widetilde{m0n}\rangle$ $(m, n \in {0,1})$ denotes the eigenstate of the system Hamiltonian $H_s$ that has the greatest overlap with the bare state $|m0n\rangle$, and $E_{|\widetilde{i0j}\rangle}$ represents the corresponding eigenenergy. As illustrated in Fig.~\ref{fig:one}(b), the energy levels of the system are represented by the three-component ket vectors $|Q_1Q_cQ_2\rangle$, corresponding to $Q_1$, the coupler, and $Q_2$, respectively. For simplicity, when focusing on the subspace defined by the transmon qubits, we use the notation $|Q_1Q_2\rangle$.

\begin{figure}
\begin{center}
\includegraphics[width=8.00cm, height=6.50cm]{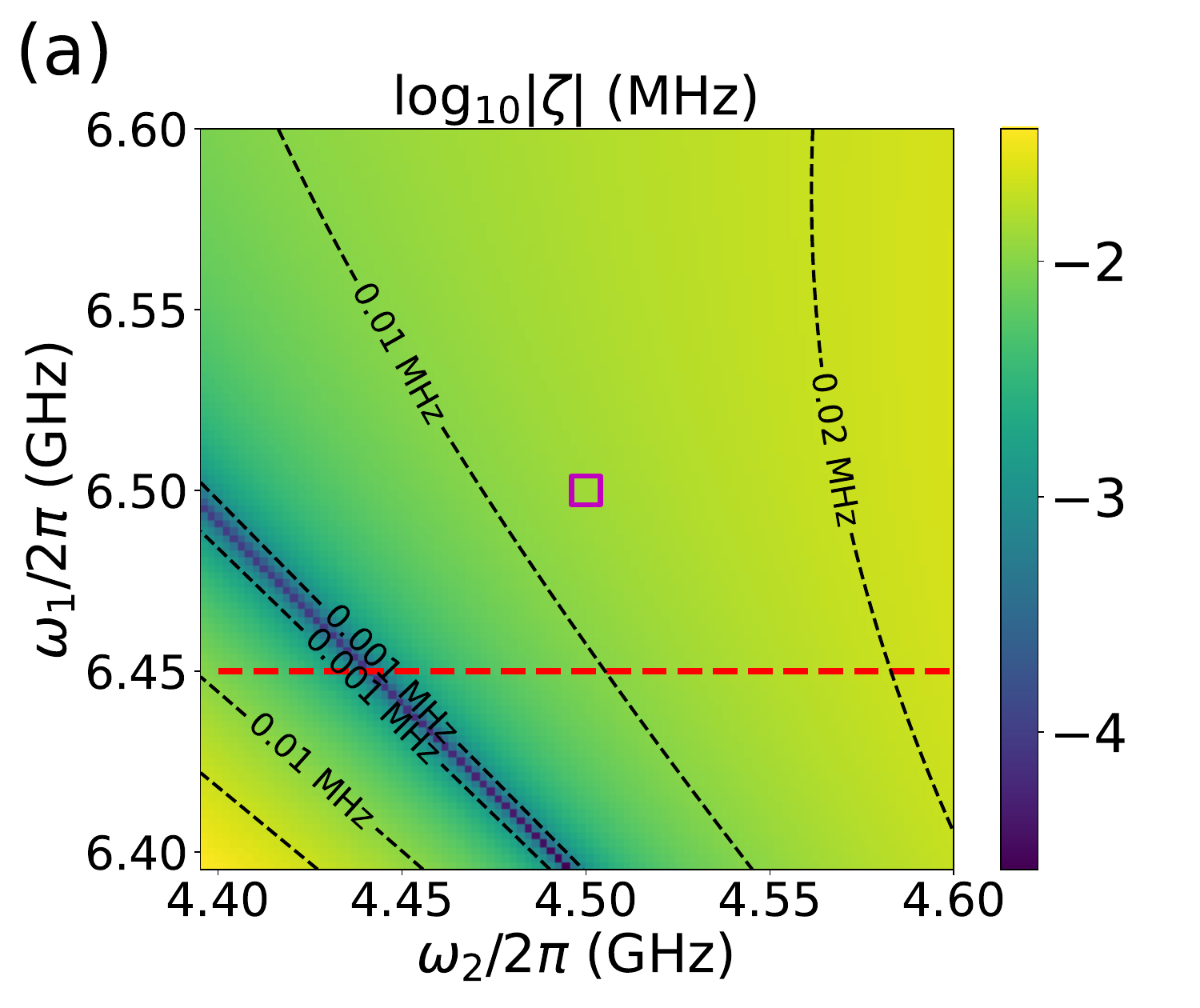}
\includegraphics[width=8.00cm, height=3.50cm]{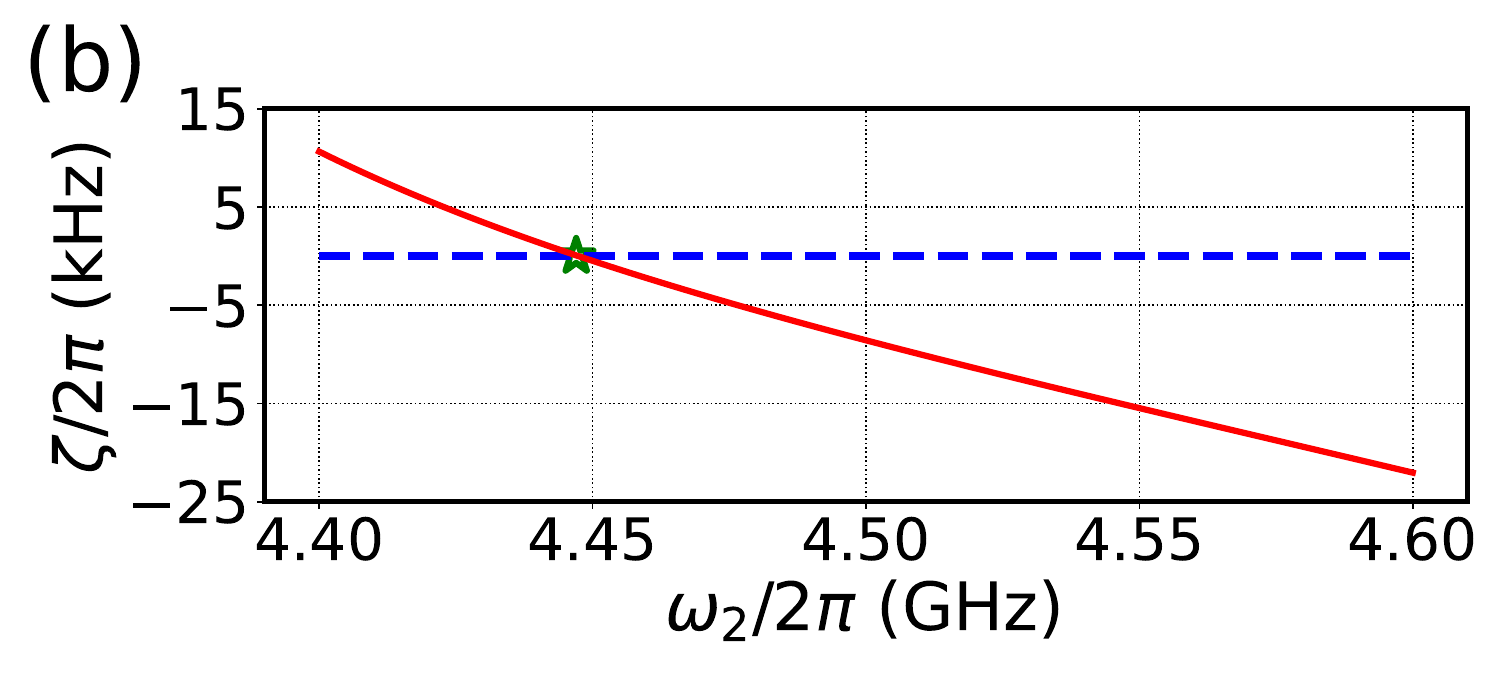}
\end{center}
\caption{(a) The ZZ interaction strength as a function of qubit frequencies $\omega_1$ and $\omega_2$, with a qubit-coupler coupling strength of $g_{1(2)c}/2\pi = 80$ MHz and a coupler frequency of $\omega_c/2\pi = 5.5$ GHz. The square marker highlights the qubit parameters used in the present work, i.e., $\omega_1/2\pi = 6.5$ GHz and $\omega_2/2\pi = 4.5$ GHz, giving rise to ZZ coupling interaction of 12.57 kHz. (b) The ZZ coupling strength vs qubit frequency $\omega_2$ with fixed qubit frequency $\omega_1/2\pi = $ 6.45 GHz, corresponding to the red cut line in (a). The blue cut curve and green star point indicate the zero value of ZZ coupling strength.} 
\label{fig:two}
\end{figure}

Before discussing our proposed microwave-activated gate scheme, we first analyze the dependence of ZZ interaction on the qubit frequencies in our proposed architecture. By maintaining a fixed coupler frequency of $\omega_c/2\pi = 5.5$ GHz and qubit-coupler interaction strength of $g_{1(2)c}/2\pi = 80$ MHz, we present the numerical results in Fig.~\ref{fig:two}(a). The ZZ coupling between qubits can be substantially reduced in the current qubit and coupler energy level configuration ($\omega_1 > \omega_c > \omega_2$ or $\omega_1 < \omega_c < \omega_2$), where ZZ coupling can be decreased to below 1 kHz, as highlighted by the dark-blue region in Fig.~\ref{fig:two}(a). Figure~\ref{fig:two}(b) illustrates the variation in ZZ coupling as the frequency of $Q_2$ is adjusted while keeping the frequency of $Q_1$ fixed. The static ZZ coupling strength can be reduced to zero, as indicated by the green star point.

This suppression occurs because ZZ coupling primarily arises from interactions between the computational state $|101\rangle$ and the non-computational states {$|200\rangle$, $|020\rangle$, $|002\rangle$}. According to fourth-order perturbation theory, the ZZ coupling strength $\zeta$ is proportional to $J^2$ \cite{PengZhaoPRAp0240372021} (see Appendix A for further details), where $J = g_{1c}g_{2c}/2(1/\Delta_1 + 1/\Delta_2)$ \cite{RJSchoelkopfNature4492007}, and $\Delta_{1(2)} = \omega_{1(2)} - \omega_c$ denotes the detuning between $Q_{1(2)}$ and the coupler. Consequently, in the current qubit-coupler energy level configuration, if the detunings $\Delta_1$ and $\Delta_2$ of the qubits are of opposite signs, the resulting qubit-qubit coupling is effectively nullified, thus minimizing the ZZ coupling. The magenta square in Fig.~\ref{fig:two}(a) indicates a ZZ coupling strength of $\zeta/2\pi = 12.57$ kHz, corresponding to the selected qubit frequencies $\omega_1/2\pi = 6.5$ GHz and $\omega_2/2\pi = 4.5$ GHz (also summarized in Table~\ref{tab:CZ_params}). This scenario will be explored further in the context of implementing a two-qubit CZ gate in the subsequent sections.

\begin{figure}
\begin{center}
\includegraphics[width=6.60cm, height=5.50cm]{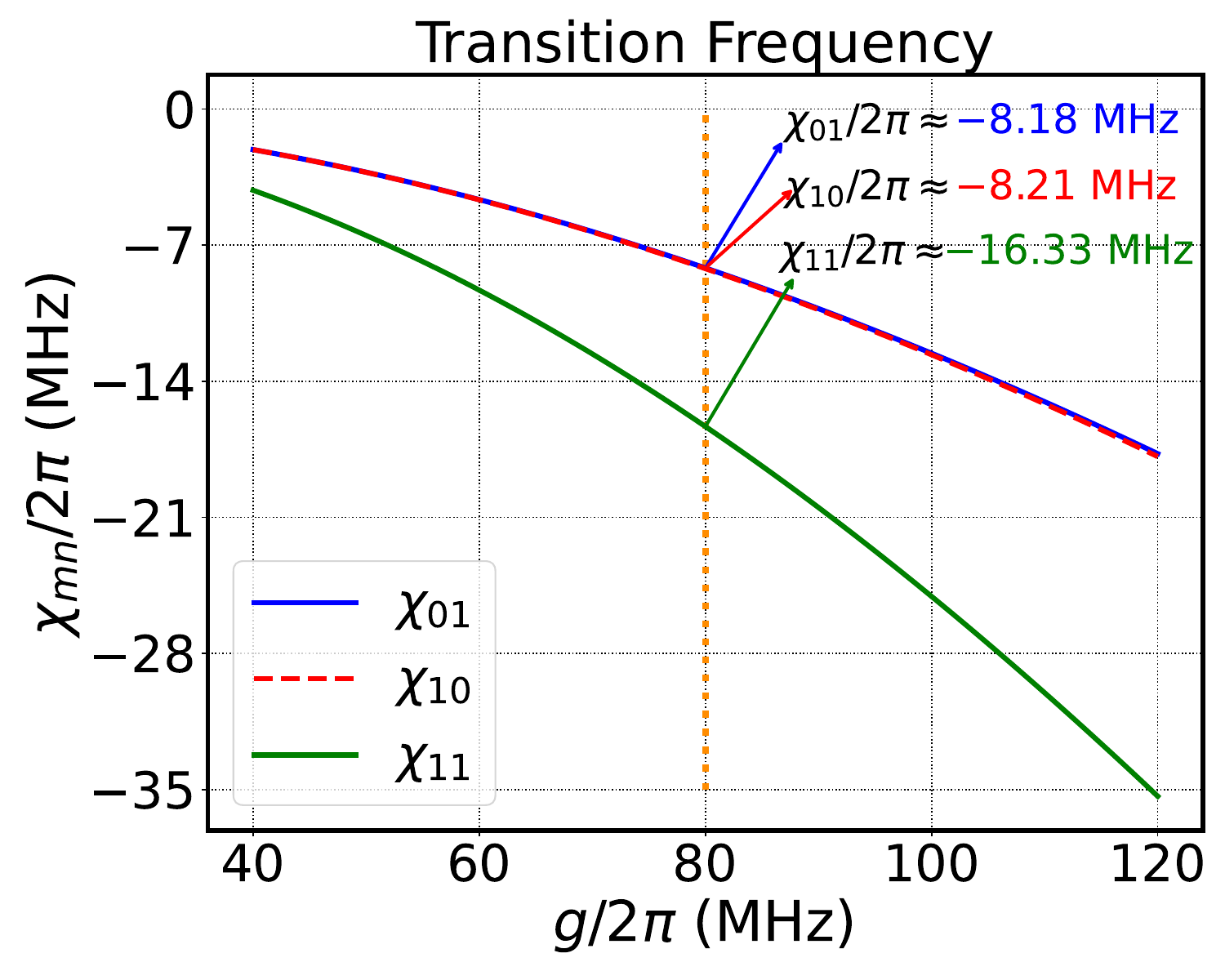}
\end{center}
\caption{Energy difference $\chi_{mn}$ = $\omega_{c,mn} - \omega_{c,00}$ $(m,n \in\{0,1\})$ between the computational basis states and their related transition states in Fig.~\ref{fig:one}(b). Here, consider the qubit-coupler couplings $g_{1c}$ and $g_{2c}$ are the same and both equal to $g$, and the other applied system parameters are presented in Table~\ref{tab:CZ_params}.}
\label{fig:three}
\end{figure}

\begin{table}[h]
\centering
\caption{System parameters for the two-qubit CZ gate.}
\label{tab:CZ_params}
\begin{tabular}{@{}lccccc@{}}
\toprule
              & Bare frequency (GHz) & Anharmonicity (MHz)   & Coupling (MHz)  \\
\hline
$Q_1$ &  $\omega_1/2\pi$ = 6.5        &          $\alpha_1/2\pi = -300$    & \multicolumn{1}{c}{\multirow{2}{*}{$g_{1c}/2\pi$ = 80}}    \\
$Q_c$     & $\omega_c/2\pi$ =  5.5        &             $\alpha_c/2\pi = -300$      &   \multicolumn{1}{c}{\multirow{2}{*}{$g_{2c}/2\pi$ = 80}}  \\
$Q_2$ &  $\omega_2/2\pi$ = 4.5        &          $\alpha_2/2\pi = -300$    &                               \\  
\toprule
\end{tabular}
\end{table}

\subsection{CZ gate scheme}

To implement a microwave-activated two-qubit CZ gate, an external microwave pulse is applied to the coupler, as described by the drive Hamiltonian: 
\begin{equation} 
\begin{aligned} 
H_d = \frac{\Omega_d(t)}{2} \left[ e^{-i(\omega_d t + \phi_0)} a^\dagger_c + e^{i(\omega_d t + \phi_0)} a_c \right], 
\end{aligned} 
\end{equation} 
where $\Omega_d(t)$ denotes the time-dependent pulse envelope, $\omega_d$ represents the drive frequency, and the initial phase $\phi_0$ is set to zero for simplicity. By applying a drive pulse to the coupler and choosing an appropriate drive frequency, an effective interaction between the wanted quantum states can be obtained. The computational states are shown in orange in Fig.~\ref{fig:one}(b), with the coupler remaining in the ground state depicted in blue. Additionally, we consider the full Hamiltonian $H_{\text{full}} = H_s + H_d$ in the laboratory frame, accounting for the rotation at drive frequency $\omega_d$ and extending beyond the rotating-wave approximation (RWA), resulting in: 
\begin{equation} 
\begin{aligned} 
H_{\text{full}} &= \sum_{l} \left[ (\omega_l - \omega_d) a^\dagger_l a_l + \frac{1}{2} \alpha_la^\dagger_l a^\dagger_l a_l a_l \right] \\ 
& + \left[ \sum_{l \neq c} g_{lc} (a^\dagger_l a_c + a^\dagger_c a_l) + \frac{\Omega_d(t)}{2}(a^\dagger_c + a_c) \right]. 
\end{aligned} 
\end{equation}

Due to the dispersive coupling between the qubits and the coupler, the coupler frequencies vary depending on the two-qubit state. To be more specific, when the qubit is in the $|m0n\rangle$ $(m, n \in \{0,1\})$ state, the transition from state $|m0n\rangle$ to $|m1n\rangle$ is $\omega_{c,mn}$. When considering the drive pulse is applied between the state $|000\rangle$ and $|010\rangle$, the transition frequency is denoted as $\omega_{c,00}$. Here, we define $\chi_{mn}$ represents the transition frequency difference relative to $\omega_{c,00}$. Using the system parameters listed in Table~\ref{tab:CZ_params}, we have numerically plotted the variation in the frequency difference as a function of the qubit-coupler coupling strength, as illustrated in Fig.~\ref{fig:three}. The data indicate that $\chi_{mn}$ increases with a stronger qubit-coupler coupling strength, with $\chi_{10}$ and $\chi_{01}$ showing nearly identical changes. In principle, one can select one of the four transitions, i.e., $|m0n\rangle \leftrightarrow |m1n\rangle$, to realize a CZ gate operation that ensures that the computational quantum states return to their initial states after one evolution cycle, while allowing the driven state $|m0n\rangle$ to accumulate a $\pi$ phase.

\begin{figure*}
\begin{center}
\includegraphics[width=18.0cm, height=5.50cm]{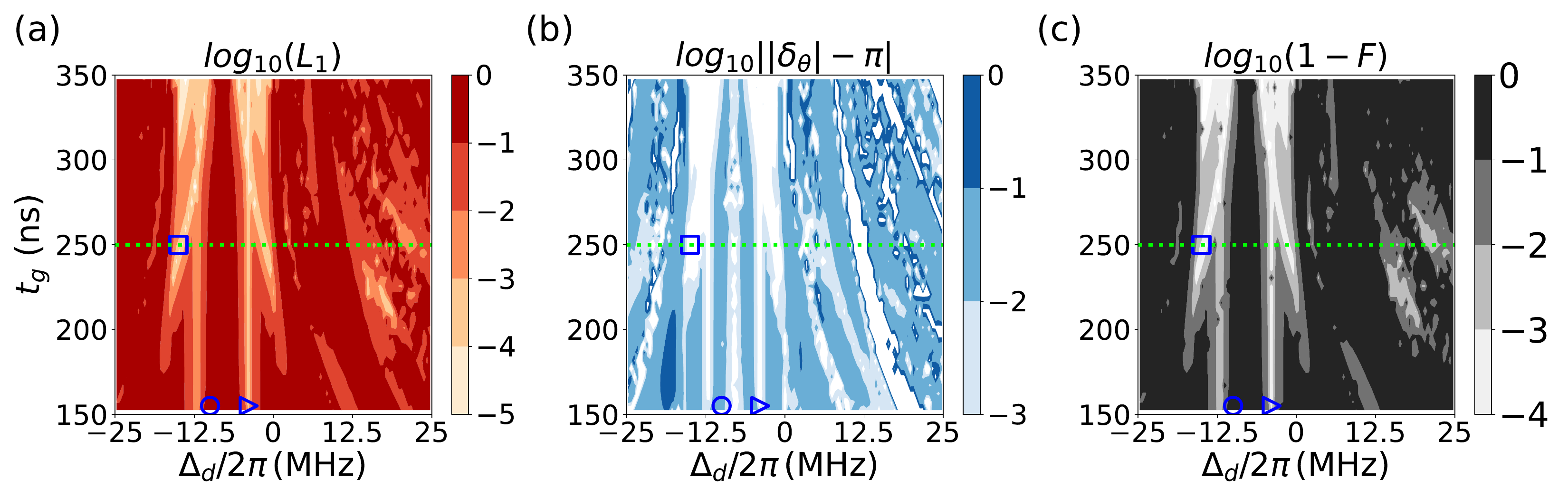}
\includegraphics[width=18.0cm, height=5.00cm]{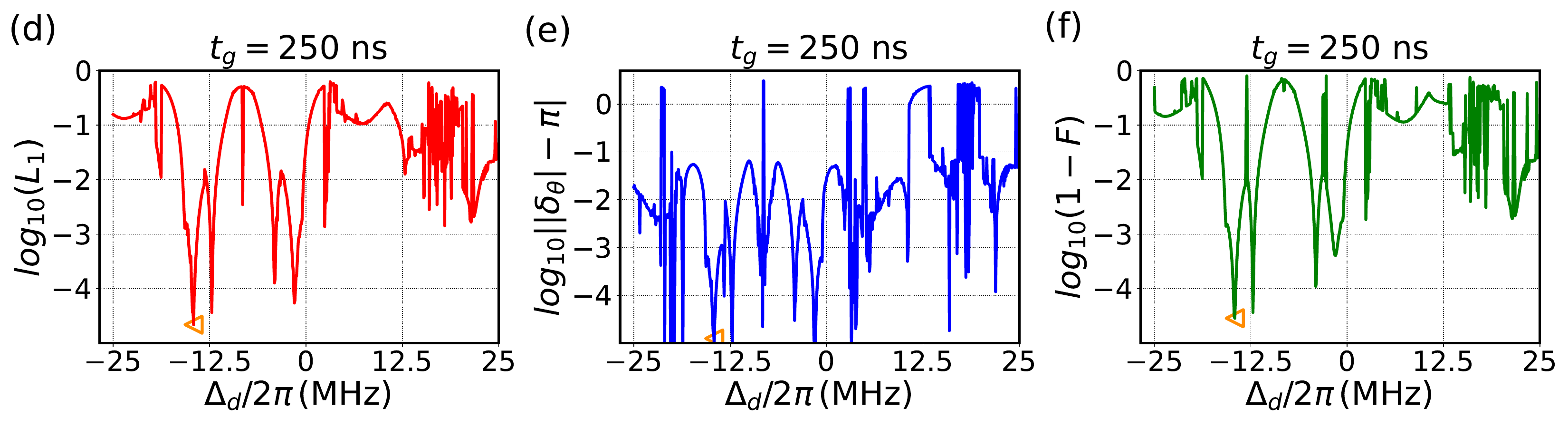}
\end{center}
\caption{(a) Leakage error $\mathcal{L}_{1}$ as a function of gate time and drive detuning is shown. (b) Variation in the phase error $||\delta_{\theta}| - \pi|$ concerning gate time and drive detuning $\Delta_d$ is presented. (c) The infidelity ($1-F$) as a function of gate time and drive detuning $\Delta_d$ is illustrated. The blue square highlights the selected gate time of 250 ns and drive detuning $\Delta_d/2\pi = -15$ MHz which leads to the gate fidelity of 0.9996 with the microwave pulse parameters $\Omega_0/2\pi \approx 8.3$ MHz, $\lambda_1 \approx 0.3395$, $\lambda_2 \approx 0.0601$. The blue circle and triangle-right points demonstrate the case of a short gate time of 150 ns. The leakage error (d), phase error (e), and infidelity (f) as functions of drive detuning with a fixed gate time of $t_g = 250$ ns are shown, corresponding to the horizontal lime cuts in (a-c). The applied system parameters are detailed in Table~\ref{tab:CZ_params}.} 
\label{fig:four}
\end{figure*}

However, the results in Fig.~\ref{fig:three} reveal that the frequency differences $\chi_{mn}$ among the four transitions are quite small, making it challenging to distinguish between them effectively. This implies that driving one transition without inducing interactions with other computational states and related states is not feasible, as shown in Fig.~\ref{fig:one}(b). Based on these results, a synchronous method can be employed for implementing the CZ gate. Considering the full Hamiltonian, with the driving pulse, the system's quantum state evolves back to the initial state with accumulated phases $\theta_{mn}$, as expressed by: 
\begin{equation} 
\begin{aligned} 
|00\rangle &\xrightarrow{H_{\text{full}}} e^{-i\theta_{00}} |00\rangle, \ |10\rangle &\xrightarrow{H_{\text{full}}} e^{-i\theta_{10}} |10\rangle, \\
|01\rangle &\xrightarrow{H_{\text{full}}} e^{-i\theta_{01}} |01\rangle, \ |11\rangle &\xrightarrow{H_{\text{full}}} e^{-i\theta_{11}} |11\rangle. 
\end{aligned} 
\end{equation} 
A two-qubit CZ gate can be realized if the accumulated phases on the four computational basis states satisfy the condition $|\theta_{11} - \theta_{01} - \theta_{10} + \theta_{00}| = \pi$.

%\begin{equation} 
%\begin{aligned} 
%\mathcal{L}_1 = 1 - \frac{\sum\limits_{m, n \in \{0,1\}}\text{Tr}[M_1 U^{\dagger}_{\text{real}}|mn\rangle\langle mn|U_{\text{real}}]}{4}, 
%\end{aligned} 
%\end{equation}

To ensure high-fidelity CZ gates, besides the conditional phase $\delta_{\theta}$ (defined as $\delta_{\theta} = \theta_{11} - \theta_{01} - \theta_{10} + \theta_{00}$), one must also consider the leakage during the gate operation. Considering the full system space is $d=64$, the computational subspace is characterized by $d_1=4$, spanning the four computational states $\{|00\rangle, |10\rangle, |01\rangle, |11\rangle\}$. The remaining $d_2=60$ represents the additional levels where leakage dynamics may occur from the $d_1$ subspace. The leakage error for a quantum system can be defined as \cite{JayMGambettaPRA970323062018, PengXuPRA1082023}:
\begin{equation} 
\begin{aligned} 
\mathcal{L}_1 = 1 - \frac{\sum_{m, n \in \{0,1\}}\text{Tr}[M_1 U^{\dagger}_{\text{real}}|mn\rangle\langle mn|U_{\text{real}}]}{4}, 
\end{aligned} 
\end{equation}
where $U_{\text{real}}$ represents the actual evolution within the computational subspace, derived from the full system Hamiltonian $H_\text{full}$, excluding decoherence effects.

To have a high-fidelity CZ gate, both the conditional phase error (i.e., $||\delta_{\theta}| - \pi|$) and the leakage $\mathcal{L}_1$ should be minimized. This can be achieved by carefully tailoring the pulse envelope and selecting appropriate pulse parameters. For simplicity, we choose a time-dependent Gaussian drive pulse,
\begin{equation} 
\begin{aligned} 
\Omega_d(t) = \Omega_0 \left[1 - \frac{1}{2} \sum_{l=1,2,3} \lambda_l \left(1 - \cos \left(\frac{t - \frac{t_f}{2}}{t_f} \cdot 2l\pi \right) \right)\right]. 
\end{aligned} 
\end{equation}
Here, $\Omega_0$ represents the peak drive amplitude, and $t_f$ denotes the total pulse duration. The parameters $\lambda_1$, $\lambda_2$, and $\lambda_3$ control the pulse shape, with the constraint $\lambda_1 + \lambda_3 = 1$ ensuring that the drive amplitude starts and ends at zero. By adjusting the shape and parameters of the drive pulse, a two-qubit CZ gate can be realized effectively.

Following the above discussions of the CZ gate error sources, i.e., including conditional phase error and leakage, we construct a cost function $\mathcal{C}_{f}$ given by
\begin{equation} 
\begin{aligned} 
\mathcal{C}_{f}(\Omega_0, \lambda_1, \lambda_2) = ||\delta_{\theta}| - \pi|^2 + \mathcal{L}_{1}. 
\end{aligned} 
\end{equation}
Here, $||\delta_{\theta}| - \pi|$ represents the phase error following the CZ gate operation. By adjusting the parameters {$\Omega_0$, $\lambda_1$, $\lambda_2$} in the Gaussian drive pulse envelope, we can use this cost function to minimize the value, thereby ensuring that $|\delta_{\theta}|$ is approximately equal to $\pi$ and the leakage error $\mathcal{L}_{1}$ is sufficiently small.

Furthermore, evaluating the accuracy of gate operations requires the calculation of gate fidelity, a crucial metric in quantum computing. To assess the performance of the implemented two-qubit CZ gate, we employ the state-averaged gate fidelity metric, as defined in \cite{PedersenPLA2007},
\begin{equation} 
\begin{aligned} 
F = \frac{[ \text{Tr}(U_{\text{ideal}}^{\dagger} U_{\text{real}}) + |\text{Tr}(U_{\text{ideal}}^{\dagger} U_{\text{real}})|^2 ]}{20}. 
\end{aligned} 
\end{equation}
Here, $U_{\text{ideal}}$ denotes the ideal CZ gate. Besides the conditional phase $|\delta_{\theta}| \approx \pi$, the single-qubit phase can be compensated by single-qubit Z gates, which can be realized virtually by updating the phase of the following microwave drive \cite{JMGambettaPRA0223302017}.

%During the system's evolution, different quantum states accumulate distinct phase shifts. To achieve the target precision of $\delta_{\theta} \approx \pi$, single-qubit phase errors can be mitigated through single-qubit gate operations, thereby enhancing the fidelity of the CZ gate. The system dynamics enable the determination of single-qubit phases using numerical optimization techniques. Corrective single-qubit phase operations \cite{GhoshPRA2013, ZahedinejadPRAp2016, BarnesPRB2017} can then be applied immediately before and after the CZ gate to rectify these phase discrepancies.

%\begin{table*}[htbp] 
   % \caption{Result.} 
    %\label{tab:result}
    %\begin{center}
        %\begin{tabular}{|l|l|l|l|l|l|}
            %\hline
            %\multicolumn{2}{|l|}{\diagbox{Dataset}{Model}} & \rule[-1ex]{0pt}{3.5ex}  VGG-16 & GoogLeNet & ResNet-50\\
            %\hline
            %\rule[-1ex]{0pt}{3.5ex} \multirow{4}*{UCM}  & train set & 100 & 88.75 & 90.74  \\
            %\cline{2-5}
            %\rule[-1ex]{0pt}{3.5ex}   & val set & 90.00 & 92.35 & 93.70  \\
            %\cline{2-5}
            %\rule[-1ex]{0pt}{3.5ex}   & test set & 90.97 & 92.17 & 93.26  \\
            %\cline{2-5}
            %\rule[-1ex]{0pt}{3.5ex}   & overall & 97.02 & 97.69 & 98.41  \\
            %\hline
        %\end{tabular}
    %\end{center}
%\end{table*}

Based on the defined cost function $\mathcal{C}_{f}$, we can determine a set of parameter values $(\Omega_0, \lambda_1, \lambda_2)$ by minimizing this function. According to the optimized parameters, the drive pulse applied to the coupler can be designed to minimize the leakage errors and meanwhile ensure the conditional phase $\delta_{\theta}$ for the four quantum states evolves to $\pi$ in a short time. While it is well established that a longer gate operation time generally results in higher fidelity, the practical goal is to achieve the highest possible fidelity within the shortest possible duration.

\begin{figure}
\begin{center}
\includegraphics[width=8.70cm, height=4.5cm]{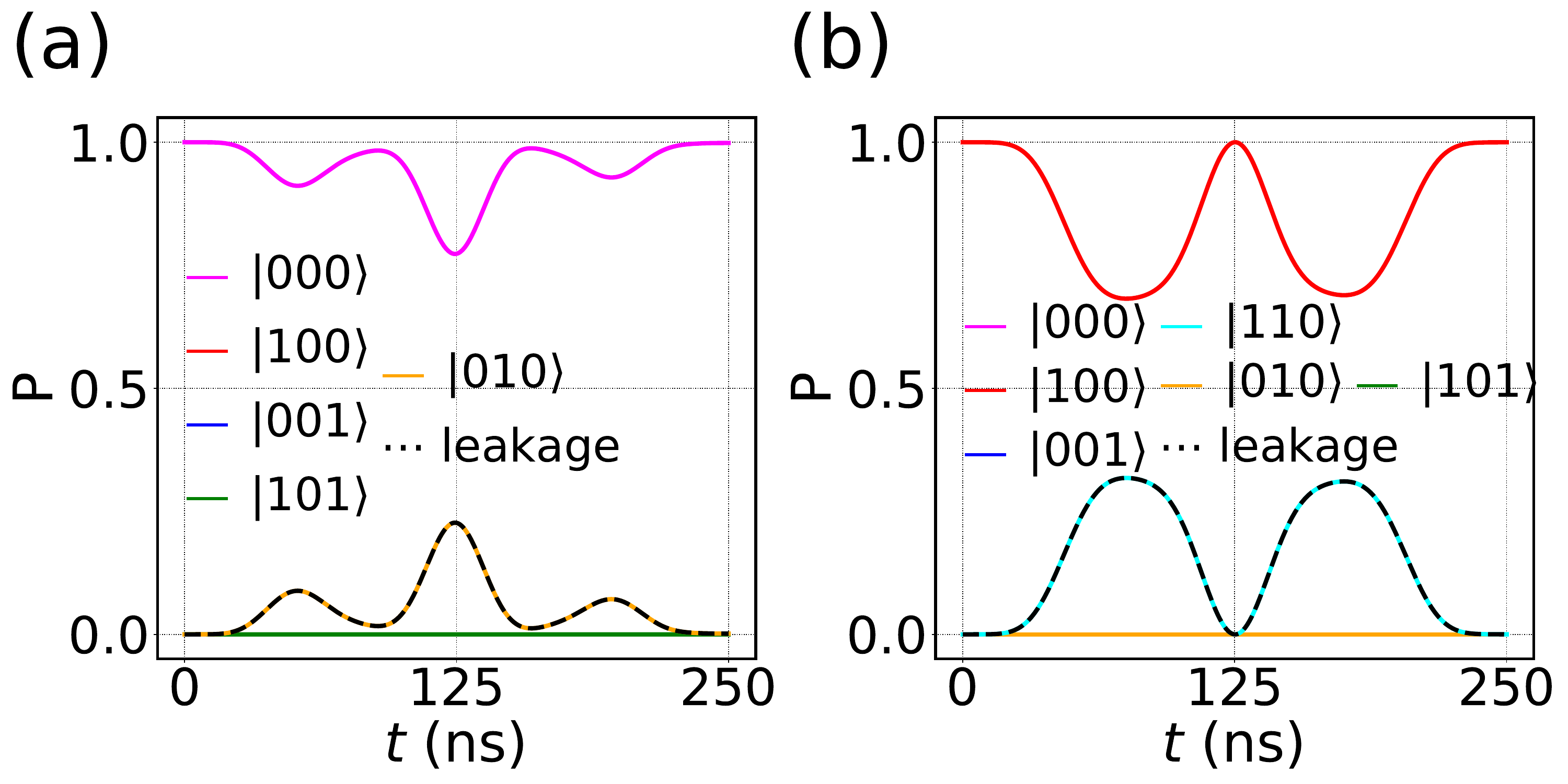}
\includegraphics[width=8.70cm, height=4.5cm]{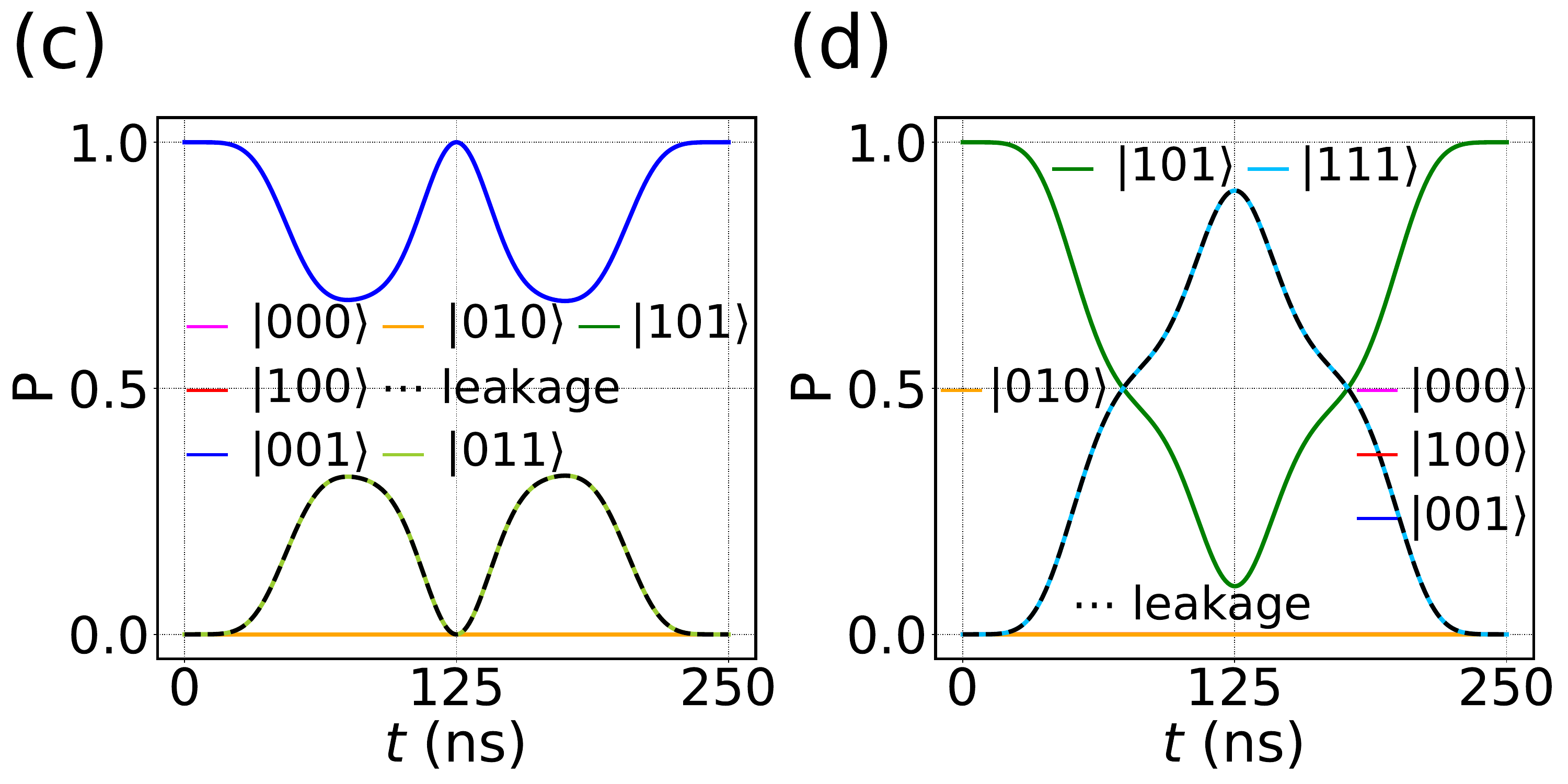}
\includegraphics[width=8.200cm, height=4.10cm]{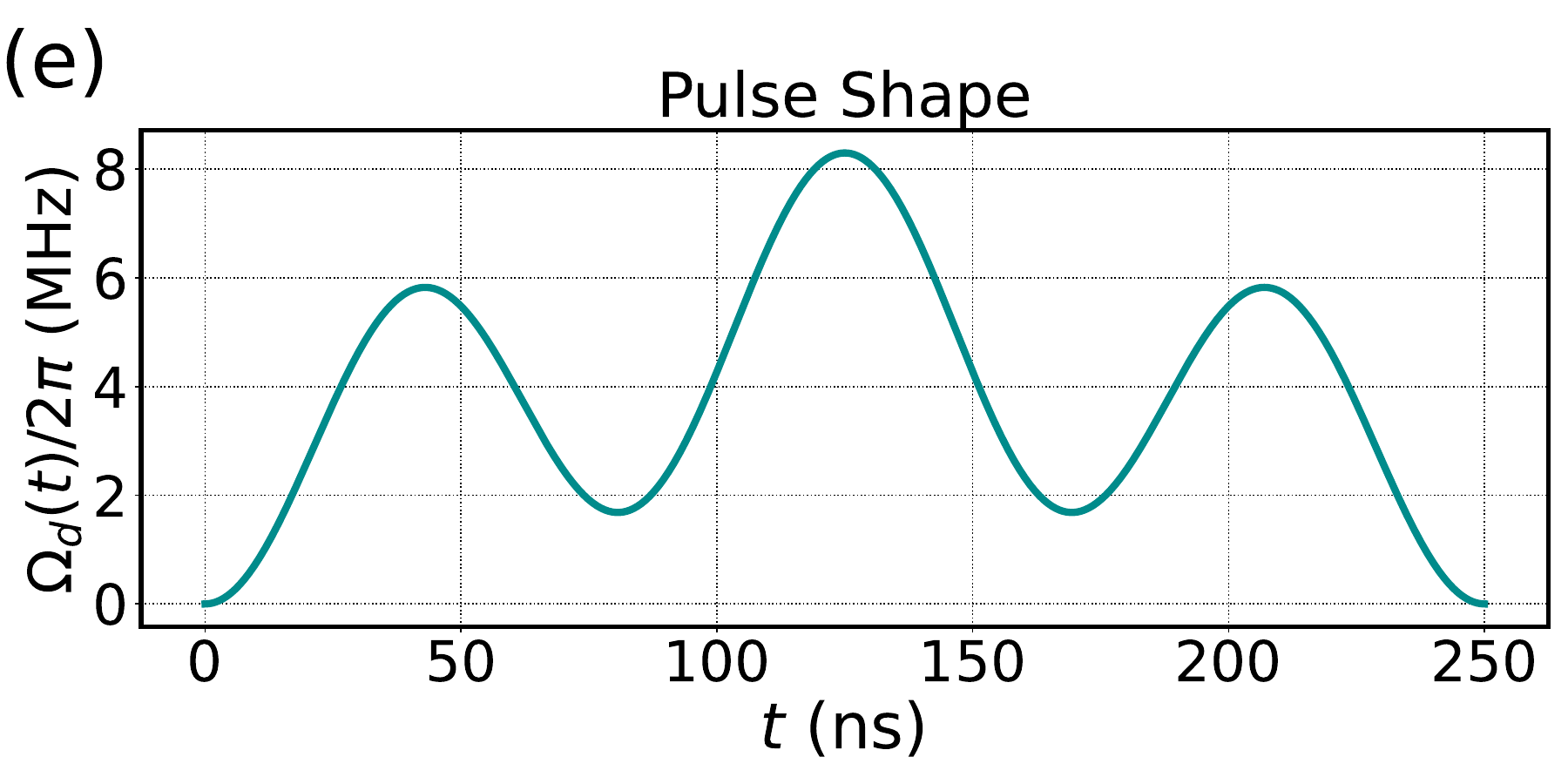}
\end{center}
\caption{The system dynamics during the gate operation with a gate time of 250 ns and a drive detuning of $-15$ MHz are indicated by the blue square in Fig.~\ref{fig:four}. In panel (a), the state $|000\rangle$ returns to its initial configuration. For initial states $|100\rangle$ (b) and $|001\rangle$ (c), the quantum states exhibit similar dynamics, with the system returning to the initial states. For these initial states, the primary leakages during the CZ operation are into $|110\rangle$ (cyan) and $|011\rangle$ (yellow-green line), respectively. Panel (d) shows the system dynamics with the initial state $|101\rangle$, where the main leakage occurs into the state $|111\rangle$ (deep-sky-blue curve) throughout the gate operation. This leakage is significantly suppressed after the gate operation. Panel (e) displays the shape of the applied drive pulse, with parameters configured as $\Omega_0/2\pi \approx 8.3$ MHz, $\lambda_1 \approx 0.3395$, $\lambda_2 \approx 0.0601$, and $t_g = 250$ ns. The values of the system parameters are provided in Table~\ref{tab:CZ_params}.}
\label{fig:five}
\end{figure}

In the two-qubit CZ gate scheme, the drive frequency is $\omega_d \approx \omega_{c,00} + \Delta_d$, where $\Delta_d$ is the drive detuning. With the specified drive pulse and system parameters, we analyze the leakage error $\mathcal{L}_{1}$, phase error $||\delta_{\theta}| - \pi|$, and gate fidelity $F$ as functions of gate time $t_g$ and drive detuning $\Delta_d$ following the gate operation, as shown in Fig.~\ref{fig:four}. Figure~\ref{fig:four}(a) shows that the leakage errors can be reduced to below $10^{-5}$. Figure~\ref{fig:four}(b) demonstrates that the phase error $||\delta_{\theta}| - \pi|$ can also be suppressed to less than $10^{-4}$. As depicted in Fig.~\ref{fig:four}(c), the fidelity of the realized CZ gate can exceed 0.9999. The blue square in Figs.~\ref{fig:four}(a-c) indicates a specific parameter set with $t_g = 250$ ns and $\Delta_d/2\pi = -15$ MHz. For this chosen point with the drive pulse parameters $\Omega_0/2\pi \approx 8.3$ MHz, $\lambda_1 \approx 0.3395$, and $\lambda_2 \approx 0.0601$, the leakage error $\mathcal{L}_{1}$, phase error $||\delta_{\theta}| - \pi|$, and fidelity $F$ are 0.0004, 0.0002 rad, and 0.9996, respectively. The blue circle and triangle-right points illustrate the case with a short gate time of 150 ns (See more results in Table~\ref{tab:CZ_results}). Furthermore, the result in Figs.~\ref{fig:four}(a-c) demonstrates that our proposed gate scheme can achieve high fidelity and minimal errors (primarily leakage and phase errors) for CZ gate operations within 150 ns or a shorter gate time. Additionally, as depicted in Figs.~\ref{fig:four}(d-f), we examine the variations in leakage, phase error, and infidelity ($1-F$) as a function of different drive detuning values, while maintaining a fixed gate time of 250 ns. As illustrated in Fig.~\ref{fig:four}(d), the leakage error can be reduced to below $10^{-4}$. Figure~\ref{fig:four}(e) demonstrates that multiple values of $\Delta_d$ result in the conditional phase equal to $\pi$. Within the 250 ns, certain drive parameter values also enable CZ gate fidelities exceeding 0.9999 with a drive detuning value $-14.57$ MHz, as shown by the orange triangle-left point in Fig.~\ref{fig:four}(f). A frequency sweep approach can be employed to identify the optimal parameters for achieving high-fidelity CZ gate operations within a short duration.

\begin{table*}[htbp] 
    \caption{The table shows the performance of controlled-Z (CZ) gates for different gate times (250 ns and 150 ns), comparing key parameters like detuning $\Delta_d$, drive strength $\Omega_0$, and optimization coefficients ($\lambda_1, \lambda_2$). It also includes metrics for leakage $\mathcal{L}_1$, conditional phase $\delta_{\theta}$, and gate fidelity $F$. The referenced symbols (\textcolor{blue}{$\square$}, \textcolor{blue}{$\bigcirc$}, and \textcolor{blue}{$\triangleright$}) in Fig.~\ref{fig:four} correspond to specific data points.} 
    \label{tab:CZ_results}
    \begin{center}
        \begin{tabular}{cc|c|c|c}
            \hline\hline
            \rule[-1ex]{0pt}{3.5ex} &\ \  \ \ &\ \ \  \textcolor{blue}{$\square$} in Fig.~\ref{fig:four} \ \ \  &\ \ \  \textcolor{blue}{$\bigcirc$} in Fig.~\ref{fig:four} \ \ \  & \ \ \  \textcolor{blue}{$\triangleright$} in Fig.~\ref{fig:four}  \ \ \  \\
            \hline
            \rule[-1ex]{0pt}{3.5ex}  &  \ \ \ \   \ \ \ \  {Gate time} \ (ns)  \ \ \ \   \ \ \ \  &\ \ \  250  \ \ \  &\ \ \  150  \ \ \  & \ \ \  150  \ \ \  \\
            \hline
            \rule[-1ex]{0pt}{3.5ex}  & \ \ \ \  $\Delta_d/2\pi \ (\text{MHz}) \ \ \ \ $ & $-15$ & $-10$ & $-3.9$  \\
            \hline
            \rule[-1ex]{0pt}{3.5ex}  &  \ \ \ \  $\Omega_0/2\pi \ (\text{MHz}) \ \ \ \  $ & $8.3$  & 9.5  &  $10.86$  \\
            \hline
            \rule[-1ex]{0pt}{3.5ex}  & $(\lambda_1, \lambda_2)$ &\ \  \ \   \ \ \ \  (0.3395, 0.0601) \ \ \ \   \ \ \ \  & $\ \ \ \   \ \ \ \  (0.0481,  0.3136) \ \ \ \   \ \ \ \  $ & $\ \ \ \   \ \ \ \  (-0.2330, 0.2517) \ \ \ \   \ \ \ \  $  \\
            \hline
            \rule[-1ex]{0pt}{3.5ex}  & $\mathcal{L}_1$ & 0.0004 & $0.3774$ & $0.0009$  \\
            \hline
            \rule[-1ex]{0pt}{3.5ex}  & $\delta_{\theta} \ (\text{rad}) $ & $3.1410$  & 3.0373 & $-3.1321$  \\
            \hline
             \rule[-1ex]{0pt}{3.5ex} & $F$ & $0.9996$ & $0.5797$ & $0.9991$  \\
            \hline\hline
        \end{tabular}
    \end{center}
\end{table*}

\section{System dynamics during CZ gate}

In the following, we examine the system dynamics during the implementation of the CZ gate, which is driven by a microwave pulse applied to the coupler. The system parameters used for the numerical simulations are detailed in Table~\ref{tab:CZ_params}. Below, we present a detailed dynamical analysis of the two-qubit CZ gate within the three-transmon system. Based on the Hamiltonian $H_{\text{full}}$, we first analyze the system dynamics for a gate operation time of 250 ns and a drive detuning of $\Delta_d / 2\pi = -15$ MHz, as indicated by the blue square in Fig.~\ref{fig:four}.

\begin{figure}
\begin{center}
\includegraphics[width=8.70cm, height=4.5cm]{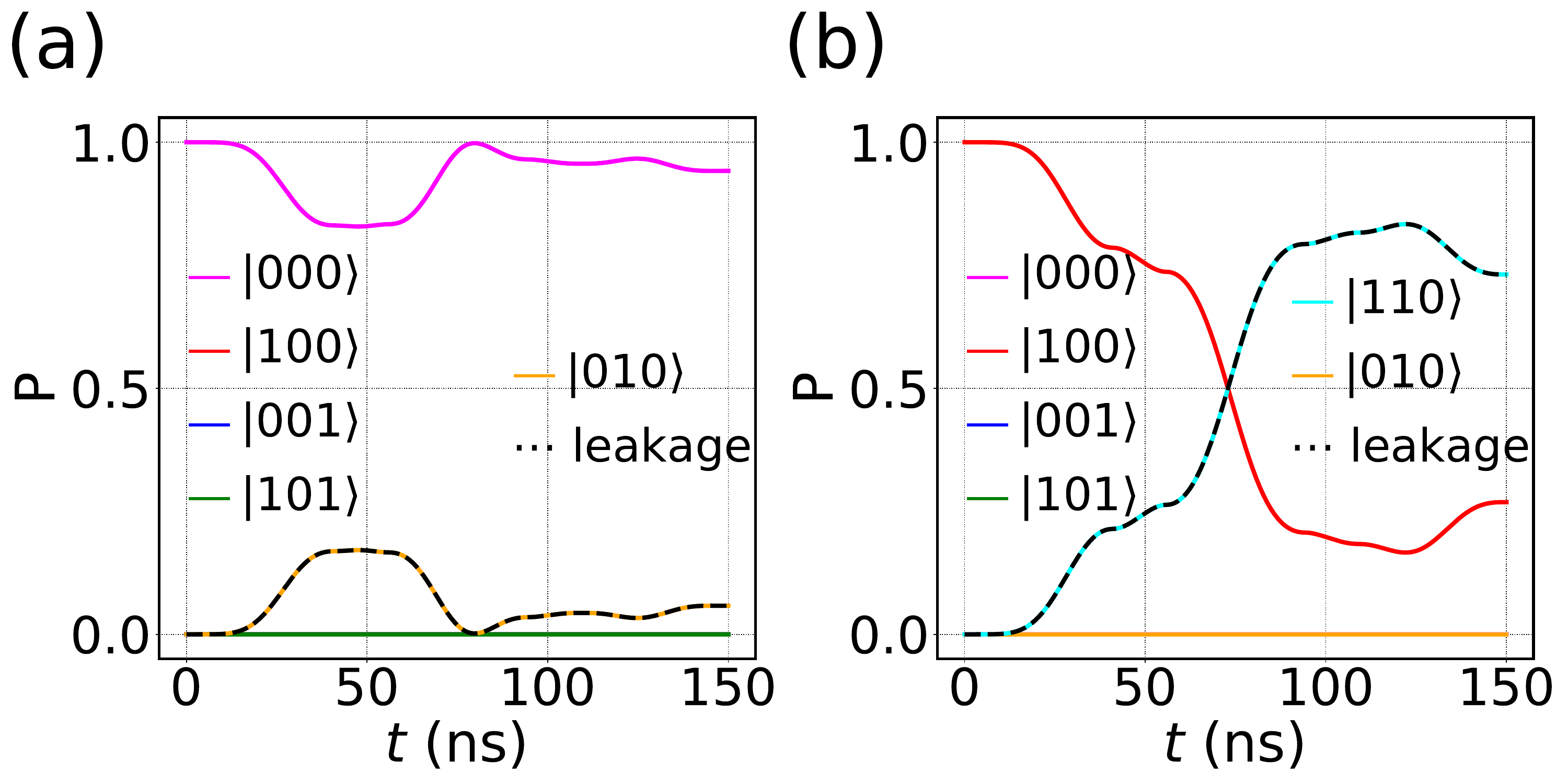}
\includegraphics[width=8.70cm, height=4.5cm]{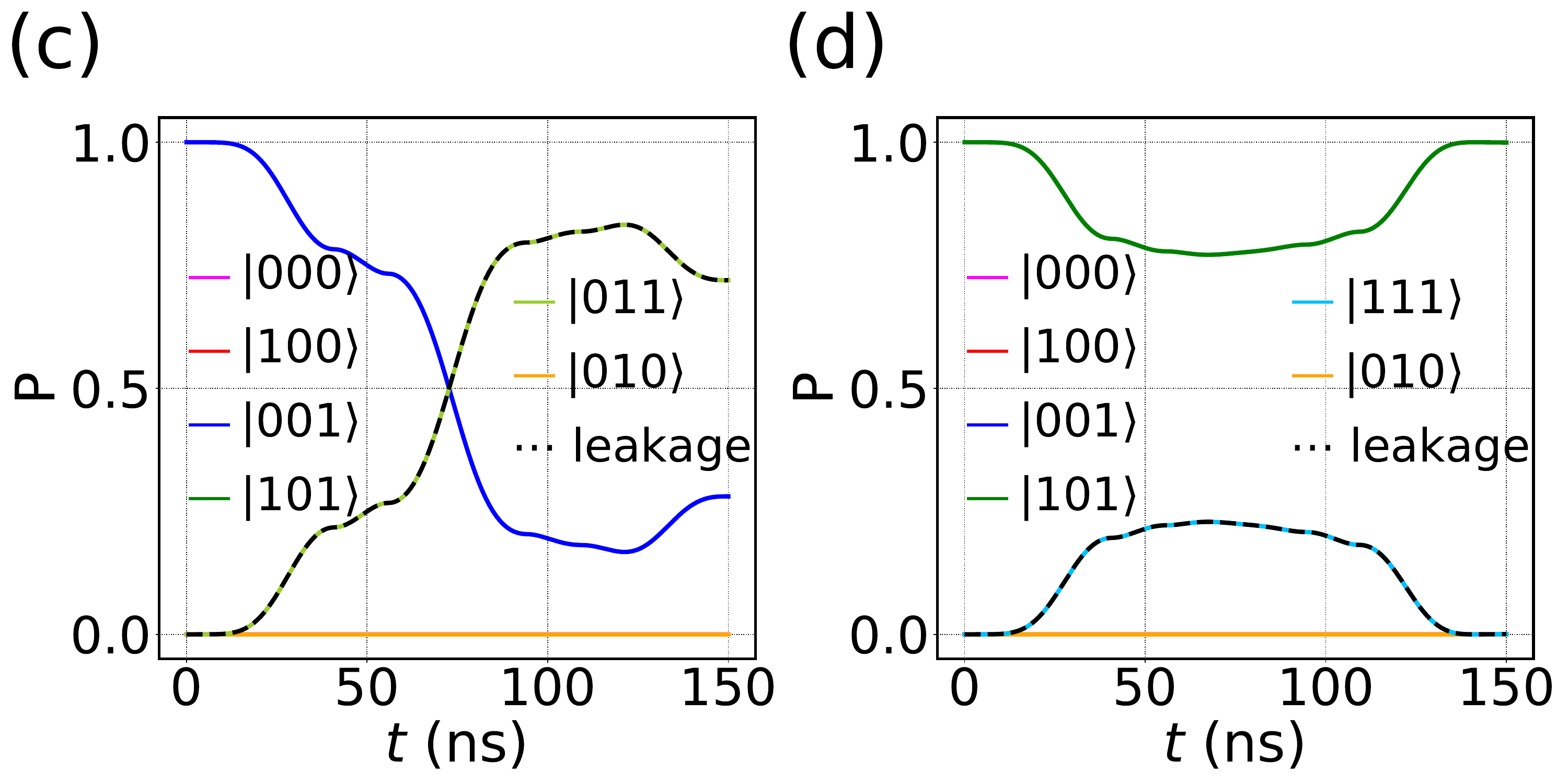}
\end{center}
\caption{Same as Fig.~\ref{fig:five}, we show the system dynamics during another short CZ gate operation. The drive pulse system parameters are set to $\Omega_0/2\pi \approx 9.5$ MHz, with pulse shape parameters $\lambda_1 \approx 0.0481$ and $\lambda_2 \approx 0.3136$. The drive detuning is $\Delta_d/2\pi = -10$ MHz, and the gate time is $t_g = 150$ ns.}
\label{fig:six}
\end{figure}

By initializing the system in four different computational states with the selected drive pulse parameters, we numerically simulate the system dynamics during the CZ gate operation, as illustrated in Fig.~\ref{fig:five}. Figure~\ref{fig:five}(a) shows that the system can evolve back to the initial state $|000\rangle$, with the primary leakage occurring in the state $|010\rangle$ throughout the gate operation. The remaining computational basis states exhibit minimal changes. Figures~\ref{fig:five}(b) and~\ref{fig:five}(c) address the cases of initial states $|100\rangle$ and $|001\rangle$, respectively. After the gate operation, the system ultimately returns to the initial states. During the gate operation, leakage primarily occurs in states $|110\rangle$ and $|011\rangle$, respectively. In Fig.~\ref{fig:five}(d), we initialize the system in the state $|101\rangle$ and observe the dynamical evolution during the gate operation. Despite the presence of an off-resonant interaction between $|101\rangle$ and $|111\rangle$, the leakage is significantly suppressed after the gate operation. To validate our theoretical model, we also numerically calculate the leakage error $\mathcal{L}_1 \approx 0.0004$, the conditional phase $\delta_{\theta} \approx 3.1410$ rad, and the fidelity $F \approx 0.9996$ with the optimized drive pulse parameters presented in Table~\ref{tab:CZ_results}. These results further confirm that our proposed gate scheme achieves a high-fidelity CZ gate with optimized drive pulse parameter values.

\begin{figure*}
\begin{center}
\includegraphics[width=18.0cm, height=5.50cm]{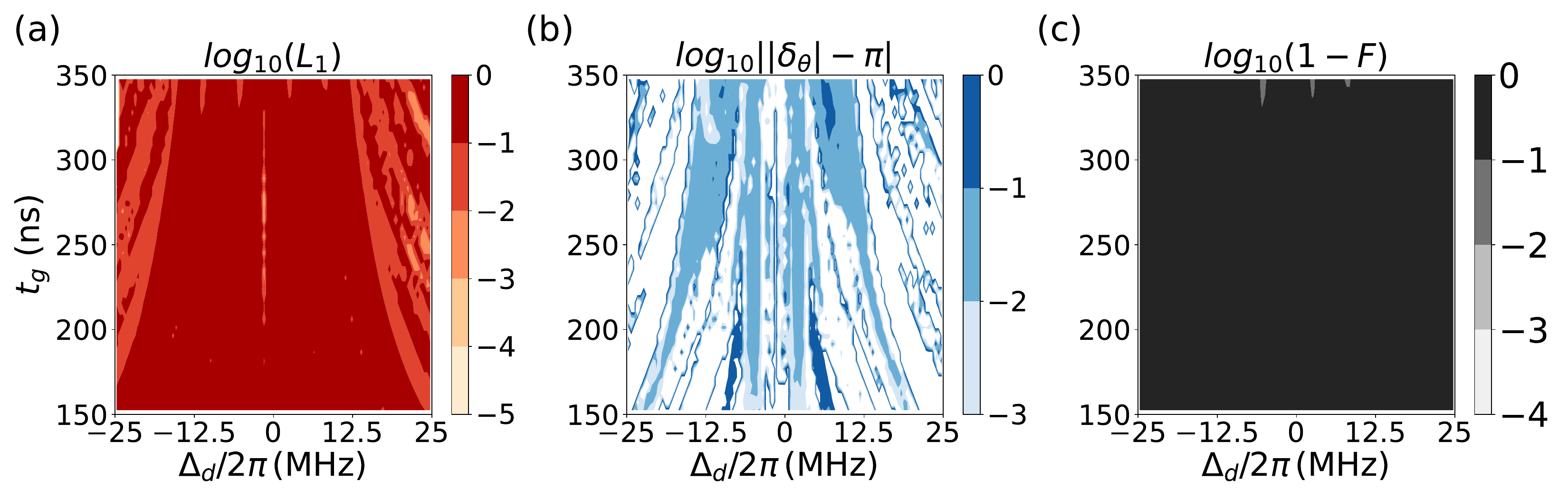}
\includegraphics[width=18.0cm, height=5.00cm]{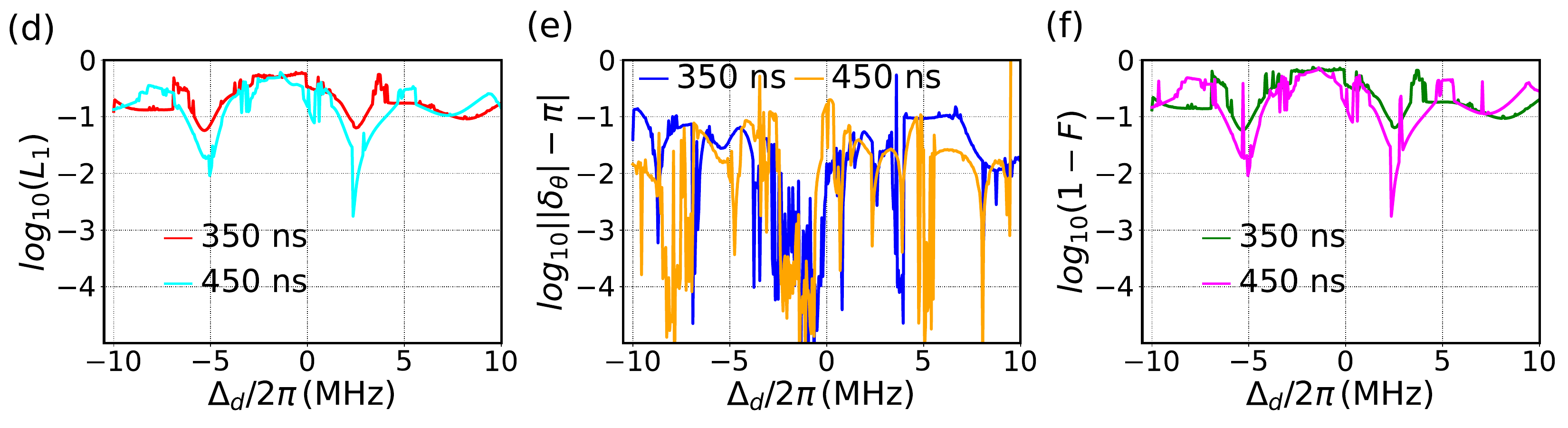}
\end{center}
\caption{The leakage, phase error, and infidelity for the CZ gate are evaluated using the system parameters outlined in Ref. \cite{YNakamuraPRL2606012023}. (a) Leakage $\mathcal{L}_{1}$ is plotted as a function of gate time and drive detuning. (b) The variation of the phase error $||\delta{\theta}| - \pi|$ is shown concerning gate time and drive detuning $\Delta_d$. (c) CZ gate infidelity ($1 - F$) is depicted as a function of gate time and drive detuning $\Delta_d$. In (d-f), we illustrate the leakage, phase error, and infidelity as a function of drive detuning for fixed gate times of 350 ns and 450 ns, using microwave pulse parameters of $\Omega_0 / 2\pi \approx 8.3$ MHz, $\lambda_1 \approx 0.3397$, $\lambda_2 \approx 0.0594$, and system parameters provided in Table~\ref{tab:CZ_PRLparams}.} 
\label{fig:seven}
\end{figure*}

Next, we examine the two-qubit CZ gate scenario for a shorter gate time. Setting the gate duration to $t_g = 150$ ns and the drive detuning to $\Delta_d / 2\pi = -10$ MHz, we optimize the drive pulse parameters to $\Omega_0 / 2\pi \approx 9.5$ MHz, $\lambda_1 \approx 0.0481$, and $\lambda_2 \approx 0.3136$. By initializing the system in the four basis states ${|000\rangle, |100\rangle, |001\rangle, |101\rangle}$, we compute the populations of these states, the first excited state of the coupler qubit, and the leakage from the computational subspace to the non-computational space states. As depicted in Fig.~\ref{fig:six}(a), the quantum state does not return to its initial state $|000\rangle$. The other three basis states remain nearly unpopulated throughout the dynamics. The populations of state $|010\rangle$ and the leakage are consistently correlated during the entire evolution, indicating that when the system is initialized in state $|000\rangle$, it entirely leaks to the non-computational state $|010\rangle$ and does not revert to its initial state post gate operation. In Figs.~\ref{fig:six}(b) and~\ref{fig:six}(c), it is evident that, for the current parameter values, the system cannot revert to the initial states $|100\rangle$ and $|001\rangle$ after the operation. The leakage primarily occurs in states $|110\rangle$ and $|011\rangle$, respectively. For an initial state of $|101\rangle$, although the system leaks to $|111\rangle$, it eventually evolves back to $|101\rangle$, as shown in Fig.~\ref{fig:six}(d). Based on the system dynamics illustrated in Fig.~\ref{fig:six}, the current parameters are insufficient for realizing the two-qubit CZ gate.

%\begin{table}[htbp] 
    %\caption{CZ gate results for the different gate time} 
    %\label{tab:result}
    %\begin{center}
       % \begin{tabular}{c|c|c|c|c|}
          %  \cline{2-5}
            %\rule[-1ex]{0pt}{3.5ex}  &\ \ \ \ &\ \ \  \textcolor{blue}{$\square$} \ \ \  &\ \ \  \textcolor{blue}{$\bigcirc$} \ \ \  & \ \ \  \textcolor{blue}{$\triangleright$}  \ \ \  \\
           % \cline{2-5}
           % \rule[-1ex]{0pt}{3.5ex}  &{\diagbox{Parameter}{Gate time}} &\ \ \  250 ns \ \ \  &\ \ \  150 ns \ \ \  & \ \ \  150 ns \ \ \  \\
            %\cline{2-5}
           % \rule[-1ex]{0pt}{3.5ex}  & $\Delta_d/2\pi \ (\text{MHz})$ & $-15$ & $-10$ & $-3.9$  \\
           % \cline{2-5}
            %\rule[-1ex]{0pt}{3.5ex}  & $\Omega_0/2\pi \ (\text{MHz})$ & $8.3$ & 9.5 & $10.86$  \\
            %\cline{2-5}
            %\rule[-1ex]{0pt}{3.5ex}  & $\lambda_1$ & 0.3395 & $0.0481$ & $-0.2330$  \\
            %\cline{2-5}
            %\rule[-1ex]{0pt}{3.5ex}  & $\lambda_2$ & 0.0601 & $ 0.3136$ & $0.2517$  \\
           % \cline{2-5}
           % \rule[-1ex]{0pt}{3.5ex}  & $\mathcal{L}_1$ & 0.0004 & $0.3774$ & $0.0009$  \\
            %\cline{2-5}
            %\rule[-1ex]{0pt}{3.5ex}  & $\delta_{\theta}$ & $3.1410$ & 3.0373& $-3.1321$  \\
            %\cline{2-5}
            %\rule[-1ex]{0pt}{3.5ex}  & $F$ & $0.9996$ & $0.5797$ & $0.9991$  \\
            %\cline{2-5}
       % \end{tabular}
   % \end{center}
%\end{table}

However, as discussed in Refs. \cite{PengXuPRA1082023, PengXuarXiv24042024} and shown in Fig.~\ref{fig:four}, we may consider adjusting the detuning parameter to $\Delta_d / 2\pi = -3.9$ MHz while maintaining the gate time at $t_g = 150$ ns for achieving the two-qubit CZ gate. Numerical calculations reveal that with optimized parameters $\Omega_0 / 2\pi \approx 10.86$ MHz, $\lambda_1 \approx -0.2329$, and $\lambda_2 \approx 0.2517$, the leakage error can be reduced to 0.0009, and the conditional phase is approximately $\delta_{\theta} \approx -3.1321$ rad. Notably, the fidelity of the realized CZ gate can reach up to 0.9991. These findings further demonstrate that our proposed gate scheme can achieve a high-fidelity two-qubit CZ operation within 150 ns.

To clearly illustrate the gate performance under different parameters, Table~\ref{tab:CZ_results} summarizes the pulse parameters and numerical simulation results discussed above. The numerical results in Figs.~\ref{fig:four}(a-c) indicate that a longer gate operation time broadens the range of optimized parameter values for achieving high-fidelity CZ gates. Additionally, for gate times shorter than 150 ns (e.g., 100 ns), optimized parameters can still be identified to ensure high-fidelity CZ gate operation.

\section{Discussions}

\begin{table}[h]
\centering
\caption{System parameters used in Ref. \cite{YNakamuraPRL2606012023}. To directly compare with our gate scheme, we omit the consideration of direct coupling between qubits.}
\label{tab:CZ_PRLparams}
\begin{tabular}{@{}lccccc@{}}
\toprule
              & Bare frequency (GHz) & Anharmonicity (MHz)   & Coupling (MHz)  \\
\hline
$Q_1$ &  $\omega_1/2\pi$ = 5.641        &          $\alpha_1/2\pi = -300$    & \multicolumn{1}{c}{\multirow{2}{*}{$g_{1c}/2\pi$ = 40}}    \\
$Q_c$     & $\omega_c/2\pi$ =  6.317        &             $\alpha_c/2\pi = -303$      &   \multicolumn{1}{c}{\multirow{2}{*}{$g_{2c}/2\pi$ = 31}}  \\
$Q_2$ &  $\omega_2/2\pi$ = 5.507        &          $\alpha_2/2\pi = -381$    &                               \\  
\toprule
\end{tabular}
\end{table}

In this section, we explore the generalization of the proposed gate scheme within the qubit-coupler system architecture utilized in Ref. \cite{YNakamuraPRL2606012023}. The system parameters are detailed in Table~\ref{tab:CZ_PRLparams}. Building on the preceding discussions, we numerically assess the leakage, phase error, and infidelity $1 - F$ as functions of gate time and drive detuning, as shown in Fig.~\ref{fig:seven}. To provide an intuitive comparison, we give Leakage, phase error, and infidelity as a function of the same region for the gate time $t_g$ and drive detuning $\Delta_d$, as illustrated in Figs.~\ref{fig:seven}(a-c). Compared to the data presented in Figs.~\ref{fig:four}(a-c), it is evident that identifying optimal values for $t_g$ and $\Delta_d$ to achieve a high-fidelity CZ gate is challenging. To streamline the calculation process and select suitable values for $t_g$ and $\Delta_d$, we fix the gate time at $t_g = 350, 450$ ns and vary the drive detuning, as shown in Figs.~\ref{fig:seven}(d-f). We find that by controlling the detuning within a range of approximately 2.5 MHz, a two-qubit CZ gate operation with minimal leakage error and high fidelity can be achieved. Subsequently, by fixing the gate time to 450 ns and drive detuning 2.5 MHz, we optimize a set of pulse shape parameters. After optimization, we determine that with drive pulse parameters $\Omega_0 / 2\pi \approx 10$ MHz, $\lambda_1 \approx -0.0178$, and $\lambda_2 \approx 0.2528$, the gate fidelity can reach 0.9947, the leakage error is about 0.0053, and the phase is approximately 3.1355 rad. Thus, the optimized drive pulse parameters can still enable a high-quality two-qubit CZ gate with different system parameter values.

\begin{figure}
\begin{center}
\includegraphics[width=8.70cm, height=4.5cm]{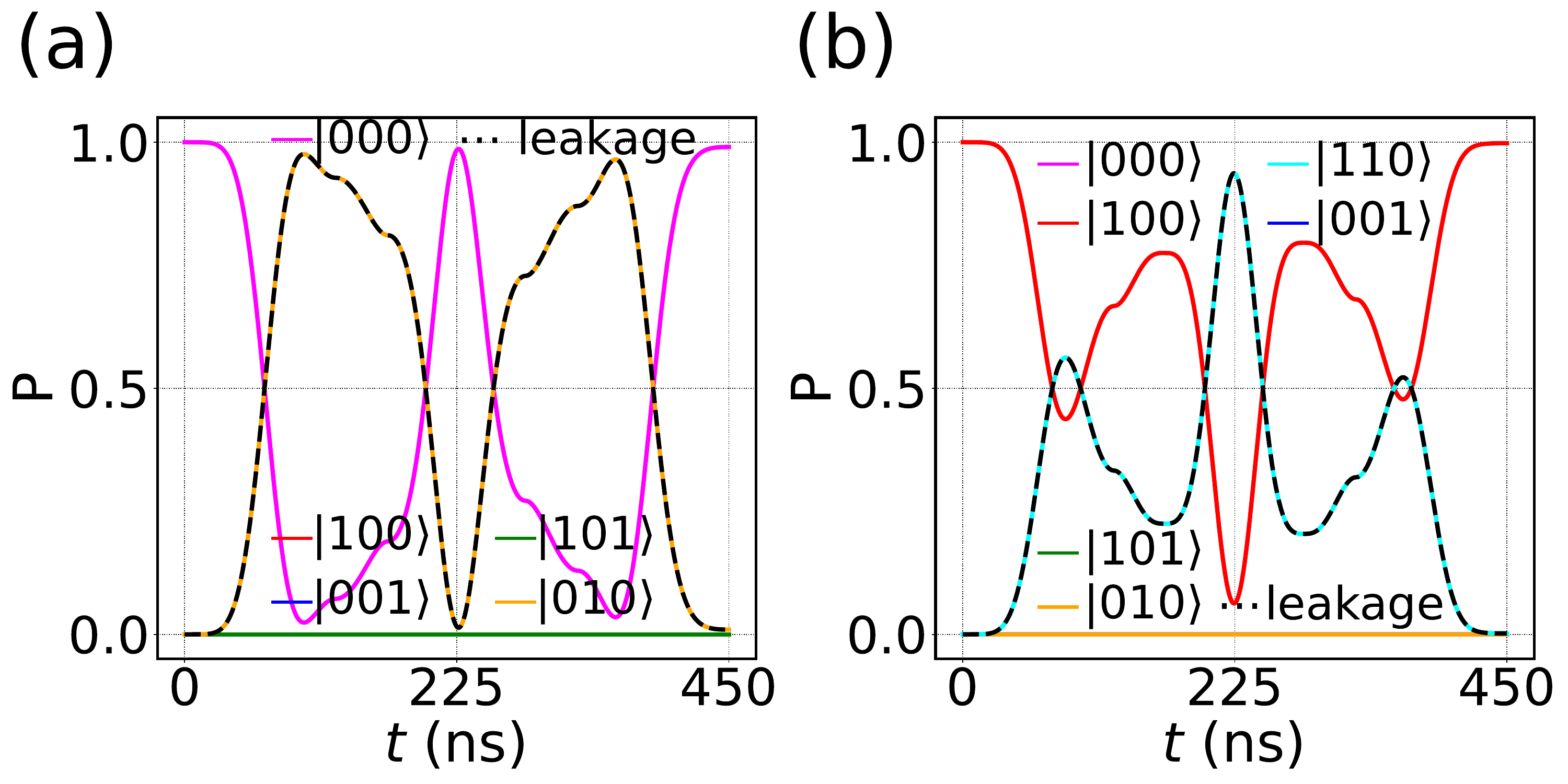}
\includegraphics[width=8.70cm, height=4.5cm]{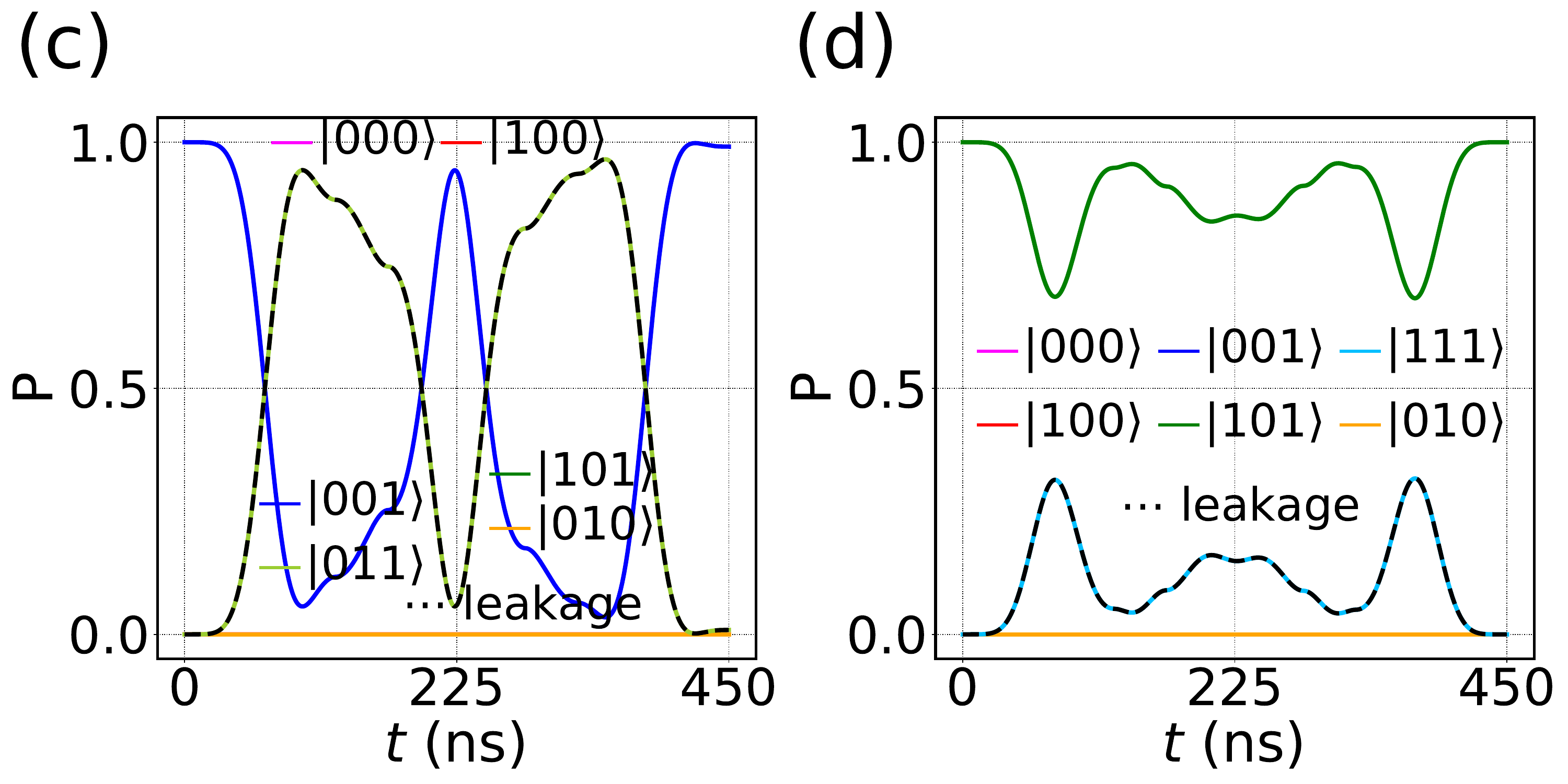}
\end{center}
\caption{To address the CZ gate operation, we numerically simulate the system dynamics, same as Fig.~\ref{fig:five}. The drive pulse system parameters are set to $\Omega_0 / 2\pi \approx 10$ MHz, with pulse shape parameters $\lambda_1 \approx -0.0178$ and $\lambda_2 \approx 0.2528$, a drive detuning of $\Delta_d / 2\pi = 2.5$ MHz, and a gate time $t_g = 450$ ns. The system parameters are detailed in Table~\ref{tab:CZ_PRLparams}.}
\label{fig:eight}
\end{figure}

To further validate this, we examine the system dynamics during the gate operation using the optimized pulse shape and system parameters, as illustrated in Fig.~\ref{fig:eight}. Figure~\ref{fig:eight}(a) shows that transitions between $|000\rangle$ and $|010\rangle$ occur due to the microwave drive. However, as the external drive diminishes to zero, the quantum system nearly returns to the initial state $|000\rangle$ with a population of 0.9902. Figures~\ref{fig:eight}(b) and~\ref{fig:eight}(c) reveal that states $|100\rangle$ and $|001\rangle$ also experience leakage during the gate operation, while states $|110\rangle$ and $|011\rangle$ exhibit complete leakage. After the gate operation, the system evolves back to the initial states $|100\rangle$ and $|001\rangle$ with populations of 0.9980 and 0.9910, respectively. Figure~\ref{fig:eight}(d) depicts the system dynamics starting from state $|101\rangle$. The results demonstrate that, despite leakage to state $|111\rangle$, the system eventually returns to state $|101\rangle$ with a population of 0.9999. These findings substantiate that the proposed gate scheme applies to various superconducting quantum system parameters. By employing an external driving pulse strength of approximately 10 MHz, the gate operation time is 450 ns. Compared to the drive amplitude (about 72 MHz) used in Ref. \cite{YNakamuraPRL2606012023}, the external drive is considerably weaker, which is advantageous for experimental implementation. Additionally, our proposed gate scheme can further reduce gate operation time and maintain high fidelity by increasing the external pulse strength and optimizing the quantum system parameters.

Before leaving this section, we note here that the preceding analysis of gate performance did not account for system decoherence. With the rapid advancements in modern technology and ongoing improvements in experimental conditions, the qubit coherence is already very high. Recent studies on superconducting quantum systems have reported decoherence times \cite{PlaceNC2021, YuHaifeng2022} that significantly exceed the gate operation time. Given that the proposed quantum gate operation scheme is executed within 150 ns, the impact of system decoherence is deemed negligible for this analysis.

\section{Conclusion}

In conclusion, we have presented a novel microwave-activated two-qubit CZ gate scheme for fixed coupling and fixed-frequency superconducting quantum systems, utilizing two transmon qubits coupled to a transmon coupler. Applying a tailored microwave pulse exclusively to the fixed-frequency coupler demonstrates that a CZ gate fidelity exceeding 0.999 can be achieved within 150 ns without considering the decoherence effect. The proposed scheme effectively minimizes leakage from the computational subspace and reduces the impact of static ZZ coupling through careful selection and optimization of qubit energy-level parameters. Moreover, this approach significantly simplifies the control complexity in multi-qubit systems by reducing the impact on adjacent qubits, thereby enhancing coherence and scalability. Numerical simulations further validate the adaptability of the proposed gate scheme across different system parameters, highlighting its potential for achieving high-fidelity gate operations with reduced driving strength and shorter gate times. The proposed microwave-activated gate scheme contributes to the ongoing efforts in optimizing quantum gate operations and paves the way for more efficient and scalable quantum computing architectures.

\section*{Acknowledgments}

L. J. would like to thank Peng Zhao for many helpful discussions on this work. This work was financially supported by the National Natural Science Foundation of China (Grants No. 12475026 and No. 12075193). P. X. was provided by the National Natural Science Foundation of China (Grants No. 12105146 and No. 12175104). S. W. also acknowledges funding from the Innovation Program for Quantum Science and Technology (2021ZD0301701) and the National Key Research and Development Program of China (No. 2023YFC2205802).

%\vspace{100mm}  
%\vskip 100cm

\appendix

\section{Perturbational analysis of ZZ coupling}

For easy reference, following Ref \cite{PengZhaoPRAp0240372021}, here we derive the effective interaction between different quantum states using perturbation theory up to the fourth order. The perturbative expression for the ZZ coupling strength denoted as $\zeta$, is given by $\zeta \equiv \zeta^{(2)} + \zeta^{(3)} + \zeta^{(4)}$, where $\zeta^{(n)} \equiv E^{(n)}_{|101\rangle} - E^{(n)}_{|001\rangle} - E^{(n)}_{|100\rangle} + E^{(n)}_{|000\rangle}$ represents the $n$th-order perturbative term, with
\begin{eqnarray} 
\begin{aligned} \label{eq5} 
E^{(2)}_{s} &= \sum_{j \ne s} \frac{|V_{sj}|^2}{E_{sj}}, \\
E^{(3)}_{s} &= \sum_{j,k \ne s} \frac{V_{sj} V_{jk} V_{ks}}{E_{sj} E_{sk}}, \\
E^{(4)}_{s} &= \sum_{j,k,l \ne s} \frac{V_{sj} V_{jk} V_{kl} V_{ls}}{E_{sj} E_{sk} E_{sl}} - \sum_{j,k \ne s} \frac{|V_{sj}|^2 |V_{sk}|^2}{E_{sj}^2 E_{sk}}, 
\end{aligned} 
\end{eqnarray}
where $V_{sj} = \langle s | H_{\text{full}} | j \rangle$ and $E_{sj} = E^{(0)}_{s} - E^{(0)}_{j}$. After applying the aforementioned approximations, and according to Eq.~(\ref{eq5}), the expressions for $\zeta^{(2)}$, $\zeta^{(3)}$, and $\zeta^{(4)}$ are as follows \cite{RJSchoelkopfNature4602009, PZhaoXuPRL1252020}:
\begin{equation} 
\begin{aligned} \label{eq6} 
\zeta^{(2)} = \zeta^{(3)} = 0, 
\end{aligned} 
\end{equation}

\begin{eqnarray} 
\begin{aligned} \label{eq8} 
\zeta^{(4)} &= 2g_{1c}^2 g_{2c}^2 \left[ \frac{1}{\Delta_{1}^2 (\Delta_{12} - \alpha_{2})} - \frac{1}{\Delta_{2}^2 (\Delta_{12} + \alpha_{1})} \right. \\ 
&\left. + \frac{1}{\Delta_{1} + \Delta_{2} - \alpha_{c}} \left( \frac{1}{\Delta_{1}} + \frac{1}{\Delta_{2}} \right)^2 \right]. 
\end{aligned} 
\end{eqnarray}
Here $\Delta = \omega_1 - \omega_2$ denotes the frequency difference between the two qubits.

\end{document}